\documentclass[preprints,article,moreauthors,accept]{mdpi} 

\usepackage{dcolumn}
\newcolumntype{d}{D{.}{.}{-1}}
\newcolumntype{f}[1]{D{.}{.}{#1}}
\newcolumntype{.}{D{.}{.}{-2}}

\firstpage{1} 
\pubvolume{1}
\issuenum{1}
\articlenumber{0}
\pubyear{2026}
\copyrightyear{2026}
\datereceived{ } 
\daterevised{ } 
\dateaccepted{ } 
\datepublished{ } 

\Title{Test of a fluorescence x-ray source for the calibration of NewAthena X-IFU}

\Author{Louis Duval $^{1,2}$\orcidA{}, François Mernier$^{2}$\orcidB{}, Akhil Mahesh$^{2}$, Alexei Molin$^{2}$\orcidC{},  Baptiste Sigal$^{2,3}$\orcidD{}, Sophie Beaumont$^{3}$, Edoardo Cucchetti$^{3}$\orcidE{} and François Pajot$^{**}$\orcidF{}}

\AuthorNames{Louis Duval, François Mernier, Akhil Mahesh, Alexei Molin, Baptiste Sigal, Sophie Beaumont, Edoardo Cucchetti and François Pajot}

\address{%
$^{1}$ \quad Dr. Karl-Remeis Observatory and Erlangen Centre for Astroparticle Physics, Friedrich-Alexander-Universität Erlangen-Nürnberg, Sternwartstr. 7, 96049 Bamberg, Germany;\\
$^{2}$ \quad IRAP, Université de Toulouse, CNRS, 9 Av. du Colonel Roche, 31400 Toulouse, France; \\
$^{3}$ \quad Centre National d'Études Spatiales, 18 Av. Edouard Belin, 31400 Toulouse, France. }

\corres{Correspondence: louis.duval@fau.de; ** francois.pajot@utoulouse.fr.}

\abstract{X-IFU is the high resolution x-ray imaging spectrometer of the NewAthena mission. X-IFU's goal is to provide high resolution spectroscopy ({4\,eV} full width at half maximum up to {7\,keV}) in the {0.2-12\,keV} energy range, over a 4\,arcmin equivalent diameter field of view thanks to an array of 1504 transition edge sensors.
To reach the energy scale accuracy goal of {0.5\,eV} up to {7\,keV}, a highly controlled ground calibration strategy is deployed.
Here, we present the commissioning of an x-ray fluorescence source using an SDD. This measurement ensured that the flux of iron {K$\alpha$} is sufficient to be observed on the microcalorimeter test bench, with a count rate of around 0.5\,counts/s/px.}


\begin{document}
\section{Introduction}

NewAthena is a large class mission from ESA, dedicated to the investigation of the hot and energetic universe such as the formation and evolution of large scale structures, as well as the growth and shape of black holes. The NewAthena x-ray telescope is a {12\,m} focal length Silicon Pore Optics mirror, made to host two instruments at its focal plane: the X-ray Integral Field Unit (X-IFU) and the Wide Field Imager (WFI)\cite{nbbf2013,mabb2021}. The X-IFU is a microcalorimeter-based x-ray imaging spectrometer\cite{bahp2013,pbca2025} with an array of 1504 transition edge sensors (TES) cooled down to {55\,mK} that aim to reach a resolution of {4\,eV} up to {7\,keV}. The instrument is optimized for photons in the energy range of 0.2\,-\,12\,keV. The energy scale 1-$\sigma$ uncertainty goal is to be below 0.65\,eV up to 7\,keV. 

\subsection{Energy scale calibration plan}

The energy scale is the function which assigns the estimated pulse height of a detected event to the energy of its corresponding photon\cite{eabb2018,labb2018,cepd2018,swab2023}. The accurate determination of this function is the key of the energy calibration of the detector. It is approximated by a high (8+) order polynomial fit to points of known energy. The energy scale is constrained by observing a set of known x-ray lines covering the entire energy band of the instrument. This step requires a careful ground calibration, to map the energy scale across its set point parameter space for a robust in-flight approximation. 
The energy scale is sensitive to the operating conditions of the detector, such as the thermal bath temperature or the radiative load and the magnetic field onto the detector\cite{pcef2016,cepd2018}. 

To do so, we need a source able to illuminate the detector array with several x-ray lines with a sufficient count rate, because the operating time of the flight model will be costly, and in order to operate within a single cold cycle of the cooling chain. For the fluorescence x-ray K-shell complex, A precision of $100$\,meV per individual complex accuracy is required to reach the total expected uncertainty of under 0.15\,eV on the energy scale calibration on the 0.2\,-\,12\,keV range. To get an estimation of the line centroid to a statistical accuracy of {$\sim$100\,meV} and sample the shape, 2500 to 5000 counts are typically needed\cite{mbap2026}.

\subsection{The Rotating Target Source}

To provide a bright and well controlled source of known X rays, we use a fluorescence source, called the rotating target source (RTS). It features a rotating wheel which allows choosing which material shines onto the detector. Such a device has already been used for the ground calibration of the XRISM mission\cite{eabb2018}, and a version has been duplicated for the ground calibration of X-IFU. With a collection of solid material across the light solid elements (magnesium up to germanium for the K$\alpha$ transitions), the RTS will be used for the energy scale calibration and will allow a complete coverage of the X-IFU energy range on the ground.

\subsection{IRAP Test bench}

To test part of the X-IFU readout electronics and commission the different calibration devices, a cryogenic test bench has been developed in Toulouse at IRAP\cite{bpra2022}. This test bench includes a complete readout chain of transition edge sensors in a commercial cryostat from Entropy Gmbh. The cryostat is cooled down with a 2-stage pulse tube cryocooler and a 2-stage adiabatic demagnetization refrigerator. These four stages are cooled down to {50\,K}, {3\,K}, {500\,mK} and {50\,mK}. This test bench enables the test of the detection chain, as well as the calibration scheme of the engineering model of X-IFU. 

\begin{figure}[]
\isPreprints{\centering}{} 
\includegraphics[width=.9\linewidth]{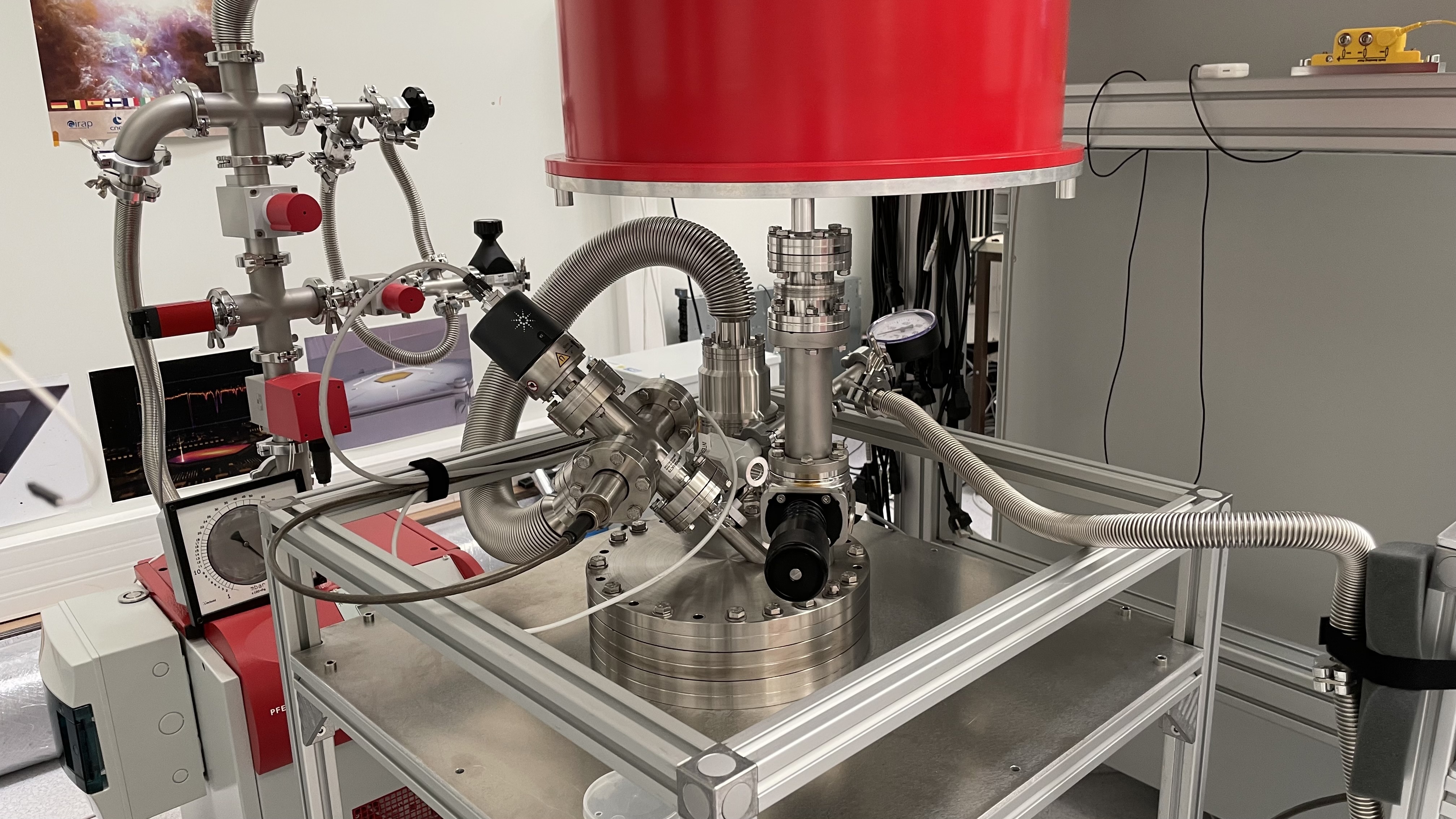}
\caption{The IRAP test bench, with the RTS placed beneath the TES sensors.\label{fig:elsa_testbench}}
\end{figure}

We report the test of the RTS with a silicon drift detector (SDD) and the first spectrum taken with the TES readout chain installed within the IRAP test bench. A picture of the RTS installed under the test microcalorimeter, is shown in Fig.~\ref{fig:elsa_testbench}.

\section{Calibration using the RTS}

Built on a similar design as the one developed at SRON, from an original design from NASA/Goddard, the IRAP RTS features a commercial soft x-ray electron impact light source from McPherson Inc. This device shines a continuum of X rays produced by an electron beam hitting a metal anode, though bremsstrahlung and impact excitation. Those X rays illuminate the rotating wheel to induce fluorescence from the selected material. The target is chosen with a rotating wheel, featuring 20 holes, 19 of which are filled with a metallic target. The list of the targets is shown in Tab.~\ref{tab:energy_table} with their respective K$\alpha$ and L$\alpha$ transition energy. Note that this range largely covers the 0.2\,-12\,keV energy range of the detector. 

\begin{table}\centering
    \begin{tabular}{c c . .}
        \hline 
        Element & Z & \multicolumn{1}{c}{L$\alpha_1$ transition energy (eV)} & \multicolumn{1}{c}{K$\alpha_1$ transition energy (eV)}  \\\hline\hline 
  Mg     &  12  &            &   1253.60  \\
  Al     &  13  &            &   1486.70  \\
  Si     &  14  &            &   1739.98  \\
  Sc     &  21  &   $395.4$     &   4090.6   \\
  Ti     &  22  &   $452.2$     &   4510.84  \\
  V      &  23  &   $511.3$     &   4952.20  \\
  Cr     &  24  &   $572.8$     &   5414.72  \\
  Mn     &  25  &   $637.4$     &   5898.75  \\
  Fe     &  26  &   $705.0$     &   6403.84  \\
  Co     &  27  &   $776.2$     &   6930.32  \\
  Ni     &  28  &   $851.5$     &   7478.15  \\
  Cu     &  29  &   $929.7$     &   8047.78  \\
  Zn     &  30  &   $1011.7$    &   8638.86  \\
  Ge     &  32  &   $1188.00$   &   9886.42  \\
  Y      &  39  &   $1922.56$   &   14958.4  \\
  Zr     &  40  &   $2042.36$   &   15775.1  \\
  Nb     &  41  &   $2165.89$   &   16615.1  \\
  Sn     &  50  &   $3443.98$   &   25271.3  \\
  W      &  74  &   $8397.6$    &   59318.24 \\
            
\hline
    \end{tabular}
    \caption{List of the 19 targets mounted on the rotating wheel, with their associated L$\alpha_1$ transition and K$\alpha_1$ transition energy transition. These values are taken from the Lawrence Berkeley National Laboratory x-ray data booklet\cite{tlal2009}.}  \label{tab:energy_table}
\end{table}

\begin{figure}[h]
\includegraphics[width=\linewidth]{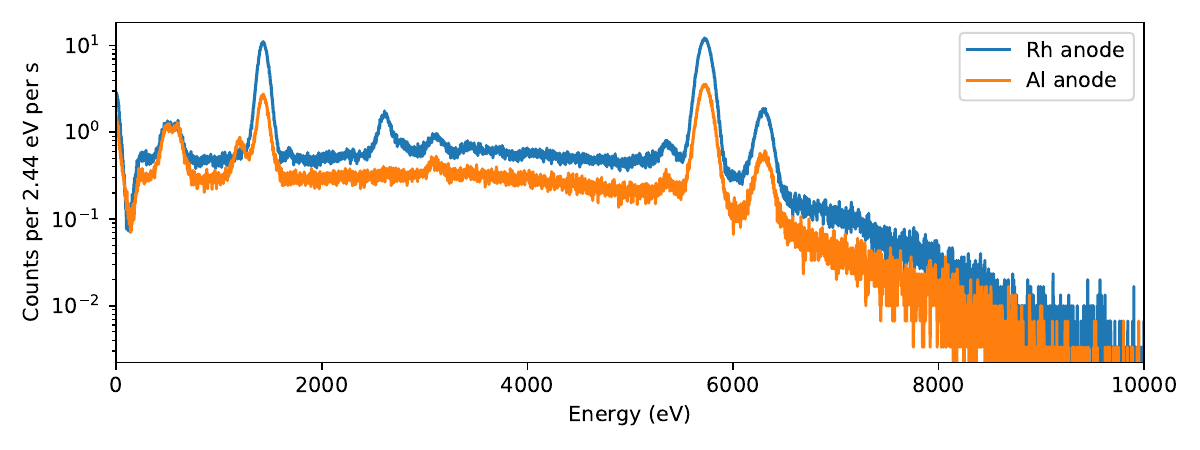}
\caption{Comparison between the manganese spectrum taken using the aluminum anode and the rhodium anode. The spectrum is roughly calibrated using a previous spectrum of magnesium. \label{fig:anode_test}}
\end{figure}

The primary x-ray source is controlled through two parameters: the voltage and the current. The voltage defines the maximum energy of emitted X rays and the current is  proportional to the count rate of emitted x-ray photons. In order to excite the lines through photoexcitation, it is mandatory that the incoming photon is of higher energy than the fluorescence line we want to observe. As the power supply is limited to {10\,kV} and {3\,mA}, only the current control is relevant for our commissioning. As we want a fluorescence which is as strong as possible, we used a current of {3\,mA} in these tests. We tested two materials for the anode of the primary source: aluminum and rhodium. As shown in Fig.~\ref{fig:anode_test}, the fluorescence yield is better with the rhodium anode than the aluminum anode, for the manganese target. The spectrum is roughly calibrated within the SDD's manufacturer software, using a previous spectrum of magnesium.

\section{Measurement of the RTS flux}

\begin{figure}[h]
\includegraphics[width=\linewidth]{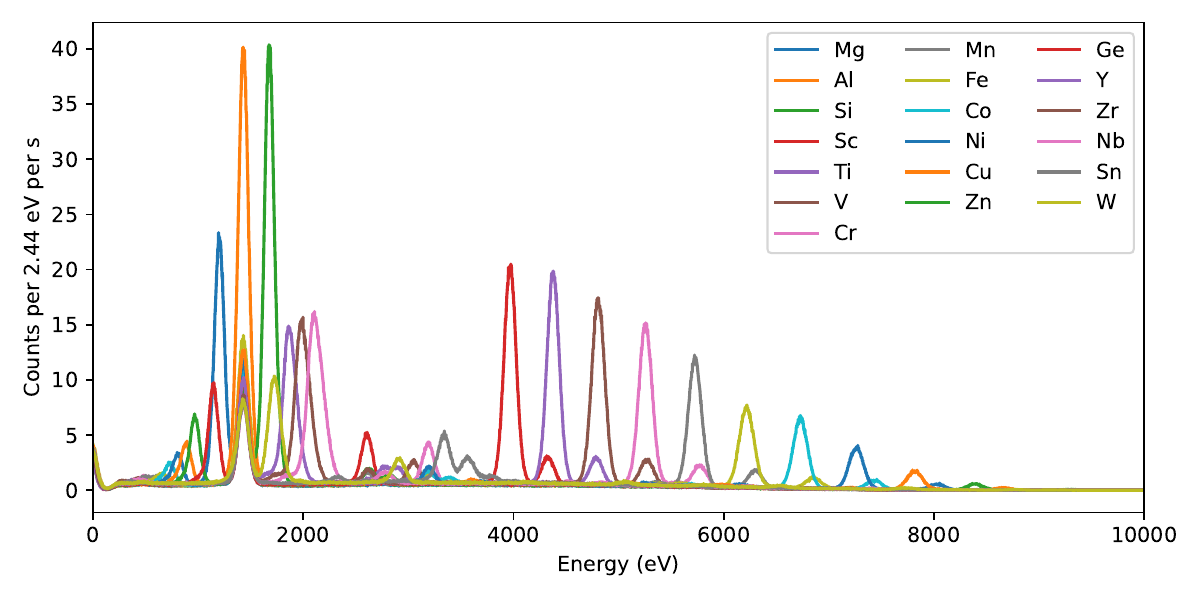}
\caption{Aggregated spectra of all the targets from the RTS, as seen by the SDD detector. The integration time is 300s. \label{fig:all_targets}}
\end{figure}

To estimate the count rate per pixel for the K$\alpha$ lines, we measured spectra with each target using the SDD. The aggregated spectra are shown in Fig.~\ref{fig:all_targets}. To get the photon number, we cut the spectrum 700\,eV below the mean K$\alpha$ energy and 800\,eV above the mean K$\beta$ energy. 
The modelling of the fluorescence lines is done through a modified Gaussian:
\begin{equation}
    \mathrm{Gaussian}(E,E_i, \mu, \sigma) = \frac{1}{\sqrt{2\pi}\sigma}\mathrm{exp}(-\frac{(x-\mu E_i)^2}{2\sigma^2}), 
\end{equation}
where $E_i$ is the mean energy of the line, $\mu$ is a calibration factor and $\sigma$ is the broadening of the line due to the detector. 
We then fit the lines using the model: 
\begin{equation}
    M(E, \mu, \sigma) = \sum_{i \in [\mathrm{K}\alpha, \mathrm{K}\beta, \mathrm{contam}]} A_i\times \mathrm{Gaussian}(E,E_i, \mu, \sigma) + bE + a,
\end{equation}
where $A_i$ is the number of photons within the line, $bE+a$ represents a background, assumed to be linear locally. We also model the contamination, such as the aluminum line, which comes from the material of the wheel, or the argon line from the detector. 

The total number of counts is displayed in Tab.~\ref{tab:photon_count_SDD}, as the sum of the K$\alpha$ and the K$\beta$ emission. This simple modelling does not grasp the full shape of the emission for some but is enough for a first estimation and a feasibility study. For elements heavier than zinc, which are elements with K-shell fluorescence energy higher than 10\,keV, no fluorescence can be induced, and do not appear in Tab.~\ref{tab:photon_count_SDD}.

\begin{figure}[h]
\includegraphics[width=\linewidth]{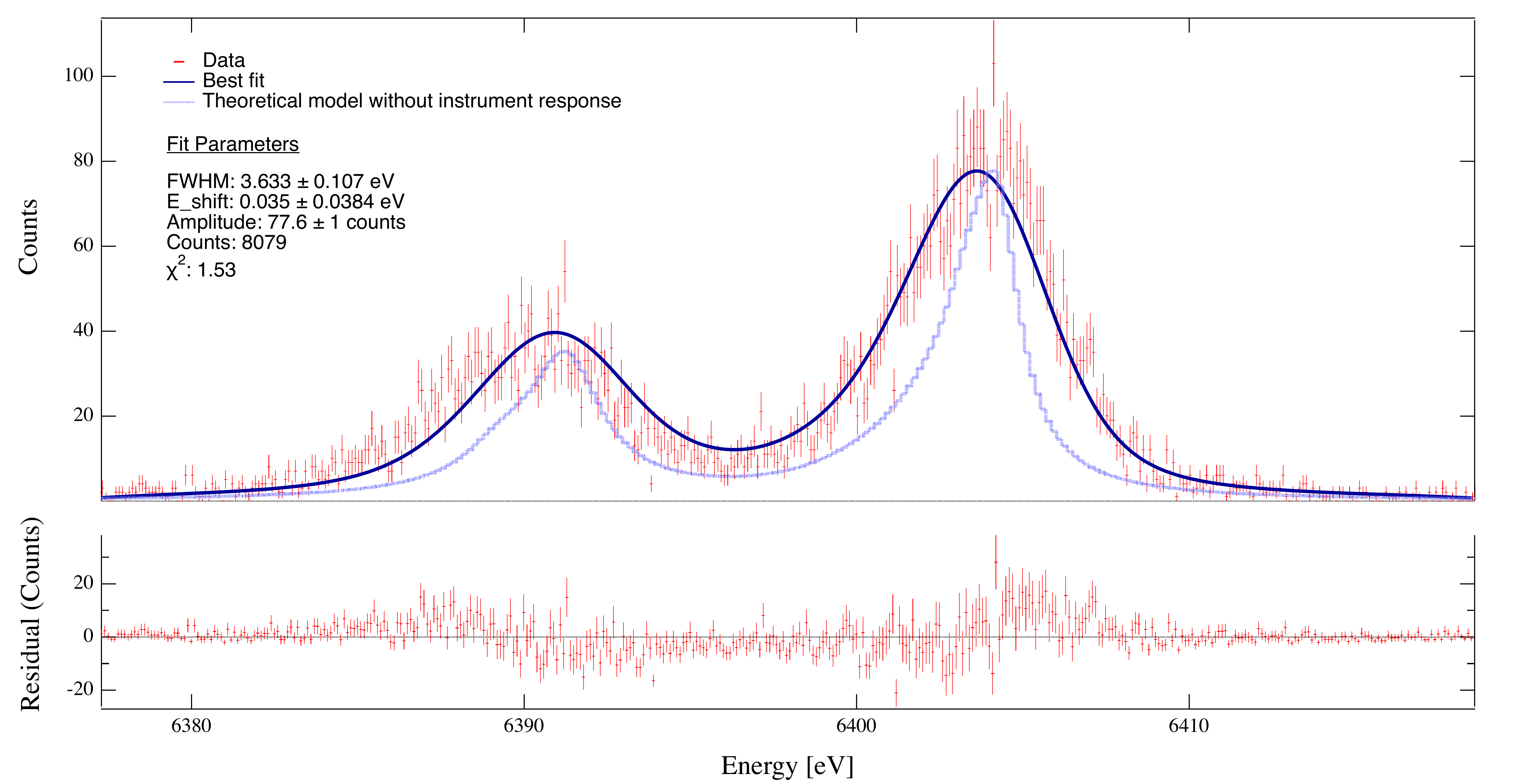}
\caption{Spectrum of iron K$\alpha_1$ and K$\alpha_2$ lines taken with the IRAP test microcalorimeter.\label{fig:microcal_spectrum}}
\end{figure}

\begin{table}[h]\centering
\begin{tabular}{cccc}
\hline
Element & $\mu$ & $\sigma$ (eV) & Count number (K$\alpha$ + K$\beta$)  \\
\hline\hline
Mg & 0.956 & 47.8 & 336922\,(580)   \\
Al & 0.959 & 48.2 & 626194\,(791)   \\
Si & 0.962 & 45.7 & 613179\,(783)   \\
Sc & 0.970 & 57.6 & 395888\,(629)   \\
Ti & 0.970 & 58.9 & 393577\,(627)   \\
V  & 0.971 & 62.3 & 363925\,(603)   \\
Cr & 0.971 & 63.8 & 323121\,(568)   \\
Mn & 0.971 & 62.6 & 252572\,(502)   \\
Fe & 0.971 & 69.4 & 174860\,(418)   \\
Co & 0.972 & 67.7 & 150362\,(387)   \\
Ni & 0.971 & 68.8 & 90573\,(300)  \\
Cu & 0.972 & 73.9 & 43137\,(207)  \\
Zn & 0.972 & 72.6 & 14047\,(118)  \\
\hline
\end{tabular}
\caption{Total number of counts for the K shell fluorescence (both K$\alpha$ and K$\beta$ emission), for a 300\,s integration time. Note that the Gaussian broadening $\sigma$ is growing is due to the degradation of the detector's resolution. $\mu$ is a calibration factor. The count number uncertainty is the Poisson uncertainty.\label{tab:photon_count_SDD}}
\end{table}
We computed the solid angle through $\Omega=A/d^2$ where A is the area of the detector and d the distance between the target and the detector. For the SDD, we have {$A=\pi(1.5\,\mathrm{mm})^2$} and $d=236.2$\,mm. Each pixel from the microcalorimeter is $\sim55$\,cm from the target with an area of $245\times245\,\mu\mathrm{m}^2$. The new count rate is then obtained by computing $C_\mathrm{microcal}=C_\mathrm{SDD}*\Omega_{microcal} /\Omega_{SDD}$, where $C_\mathrm{SDD}$ is the count rate on the SDD, $\Omega_\mathrm{SDD}$ and $\Omega_\mathrm{microcal}$ are the solid angles of, respectively, the SDD and one pixel of the microcalorimeter with respect to the target.  
With this count number estimation, we expect a count rate of roughly 1\,count/s on the microcalorimeter with an air gap of 2\,cm for the iron spectrum. This air gap is unavoidable as the microcalorimeter and the RTS are not connected for this first phase of the tests. A first iron spectrum has been taken with the microcalorimeter, achieving a mean of 0.2 to 0.5\,count/s/px. The aggregated spectrum is shown in Fig.~\ref{fig:microcal_spectrum}. The current count rate requires an exposure of several kiloseconds to reach the desired 5000\,counts for each target. 

\section{Conclusions}

We performed a flux estimation for the microcalorimeter measurement, but the soft x-ray source used for the fluorescence does not provide X rays of high enough energy to induce fluorescence above 10\,keV. This issue will be resolved by using a brighter source, with a higher energy primary source, developed in collaboration between SRON, MPIK and IRAP\cite{bgdg2026}. We also reported the first spectrum of fluorescence X rays taken with the microcalorimeter located in IRAP, with a resolution better than 4\,eV. Additional work will be carried in order to get an absolute energy calibration of the lines provided by our RTS, using an electron beam ion trap (EBIT) providing precise energy references from well known transitions of highly charged ions.

\section*{Acknowledgement}
We thank Luciano Gottardi and the team at SRON for their contribution in sharing the design, part list, and experience in the building of the RTS.

\reftitle{References}
\bibliography{./bibliography.bib}

\end{document}